\documentstyle[12pt]{article}
\newcounter{eqn}
\def\lab{\refstepcounter{eqn}\eqno(\arabic{eqn})}
\def\l#1{\lab\label{#1}}
\def\r#1{(\ref{#1})}
\def\c#1{\cite{#1}}

\begin{document}

\begin{titlepage}
\begin{flushright}
July 9, 2001\\
hep-th/0106250
\end{flushright}

\begin{centering}
\vfill
{\bf
BRST-BFV, DIRAC AND PROJECTION OPERATOR QUANTIZATIONS:
CORRESPONDENCE
OF STATES
}

\vspace{1cm}
O. Yu. Shvedov \footnote{shvedov@qs.phys.msu.su} \\
\vspace{0.3cm}
{\small {\em Sub-Dept. of Quantum Statistics and Field Theory,}}\\
{\small{\em Department of Physics, Moscow State University }}\\
{\small{\em Vorobievy gory, Moscow 119899, Russia}}

\vspace{0.7cm}

\end{centering}

{\bf Abstract}

The correspondence  between  BRST-BFV,  Dirac  and projection operator
approaches to  quantize  constrained systems is analyzed.  It is shown
that the component of the BFV wave function  with  maximal  number  of
ghosts  and antighosts in the Schrodinger representation may be viewed
as a wave function in the projection operator approach. It is shown by
using  the relationship between different quantization techniques that
the Marnelius inner product for BRST-BFV systems should be in  general
modified in order to take into account the topology of the group;  the
Giulini-Marolf group averaging prescription for the inner  product  is
obtained   from   the   BRST-BFV   method.  The  relationship  between
observables in different approaches is also found.

\vspace{0.7cm}


\vfill \vfill
\noindent

\end{titlepage}
\newpage

{\bf 1.}
There are different approaches to quantize the constrained systems.
One can use the Dirac approach and impose constraints on the states or
solve the classical constraints,  reduce the phase space and  quantize
the obtained system \c{D}.
An alternative  way  \c{BFV,KO,Henneaux} is based on extension of the
phase space with the help of ghosts and antighosts  and  imposing  the
BRST-BFV condition on the physical states.
The most  difficult  step for both approaches is to introduce an inner
product for physical states.

The purpose of this paper  is  to  analyze  the  relationship  between
states, observables  and  inner  products  in  different  quantization
methods.

{\bf 2.}  Consider  the  constrained  system which is specified by the
Hamiltonian $H$ and constraints $\Lambda_a$, $a=\overline{1,M}$, which
obey the Lie-algebra commutation relations:
$$
[\Lambda_a; \Lambda_b] = i f^c_{ab} \Lambda_c.
\l{1}
$$
In the Dirac approach,  states are specified by the  generalized  wave
functions $\Psi$ obeying the usual Schrodinger equation $i\dot{\Psi} =
H{\Psi}$ and the additional conditions of the form
$$
\Lambda_a \Psi = 0
\l{2}
$$
(for the case of the unimodular group).
It is  not  easy  to  introduce  an  inner  product  in this approach
\c{Henneaux}: one can impose additional gauge conditions and integrate
$|\Psi|^2$ over the gauge-fixed surface.

Instead of imposing the conditions \r{2} on the wave function, one can
take the constraints into account by modifying the  inner  product  of
the theory. States are specified by smooth and damping at the infinity
wave functions $\Phi$,  but the inner product  is  modified:  for  the
abelian case and constraints with continuous spectrum, the formula reads
$$
(\Phi, \prod_a 2\pi \delta(\Lambda_a) \Phi).
\l{3}
$$

Such inner products were used in \c{Marolf};  they naturally arise  in
the BFV  approach  \c{Marnelius}.  Similar formulas were introduced in
the projection operator approach \c{proj}.  Since  the  inner  product
\r{3} is  degenerate,  one  should  factorize  the  state space:  wave
functions $\Phi_1$ and $\Phi_2$  are  equivalent  if  $(\Phi,  \prod_a
2\pi\delta(\Lambda_a) (\Phi_1  -  \Phi_2))  = 0$ for all $\Phi$.  This
means that the following quantum gauge transformation
$$
\Phi \to \Phi + \Lambda_a X^a
\l{4}
$$
is admitted. One can suggest the following correspondence between wave
functions $\Psi$ and $\Phi$ \c{Marolf}:
$$
\Psi = \prod_a 2\pi\delta(\Lambda_a) \Phi.
\l{5}
$$
Indeed, the  wave function $\Psi$ satisfies the Dirac conditions \r{2}
and does  not  vary  if  $\Phi$  is  changed  by  a  zero-norm  state.
Prescription \r{3}  gives  us  an inner product for the Dirac approach
then.

{\bf 3.} To develop the BRST-BFV approach \c{BFV,KO,Henneaux,Marnelius},
it  is  necessary  to
introduce additional  degrees  of  freedom:  coordinates  and  momenta
$\lambda^a$, $\pi_a$,  $a=\overline{1,M}$, ghosts and antighosts
$C^a, \overline{C}_a$, canonically conjugated momenta
$\overline{\Pi}_a, \Pi^a$,    $a=\overline{1,M}$.    The    nontrivial
commutation relations are:
$[C^a, \overline{\Pi}_b]_+  =  [\overline{C}^a,  \Pi_b]_+ = \delta^a_b$;
$[\lambda^a, \pi_b]_-=i\delta^a_b$.
The following nilpotent BRST-BFV charge $\Omega$ is introduced:
$$
\Omega = C^a\Lambda_a - \frac{i}{2} f^a_{bc} \overline{\Pi}_a C^bC^c -
\frac{i}{2} f^a_{ba} C^b - i\pi_a\Pi^a.
\l{6}
$$
Instead of requirement \r{2},  the BRST-BFV  condition  is  imposed  on
physical states $\Upsilon$:
$$
\Omega \Upsilon = 0.
\l{7}
$$
while the gauge freedom is also allowed, the gauge transformation is:
$$
\Upsilon \to \Upsilon + \Omega X,
\l{8}
$$
so that  states  $\Upsilon$ and $e^{[\Omega,\rho]_+}\Upsilon$ are also
equivalent.

There are several prescriptions to introduce an  inner  product  for
physical states  (partial  cases  are  considered  in  \c{Razumov}  in
details). The  most  interesting  general  formula  is  the  following
\c{Marnelius}. One  considers  the  representatives of the equivalence
classes which obey the following additional conditions
$$
C^a\Upsilon = 0, \qquad \pi_a \Upsilon = 0
\l{9}
$$
which make  the  state $\Upsilon$ BRST-BFV-invariant.  Unfortunately,
the quantity  $(\Upsilon,\Upsilon)$  is  ill-defined.   However,   the
expression
$$
(\Upsilon, e^{t[\Omega,\rho]_+} \Upsilon)
\l{10}
$$
which is formally equivalent to  $(\Upsilon,\Upsilon)$  occurs  to  be
well-defined for a certain choice of the gauge fermion $\rho$,
$$\rho = - \lambda^a \overline{\Pi}_a
\l{10aa}
$$.
Let us  analyze  the  prescription  \r{10} (cf. \c{Marnelius}).
Consider  the Schrodinger
representation for the BFV wave function $\Upsilon$,
$\Upsilon = \Upsilon (q,\lambda,\Pi,\overline{\Pi})$.
The operators are rewritten then as
$$
C^a = \frac{\partial}{\partial \overline{\Pi}_a};
\qquad
\overline{C}^a = \frac{\partial}{\partial {\Pi}_a};
\qquad
\pi_a = - i\frac{\partial}{\partial\lambda^a};
\qquad
p_i = - i\frac{\partial}{\partial q^i},
$$
the left derivatives are considered here.
The inner product is indefinite \c{JMP0}
$$
(\Upsilon_1,\Upsilon_2) =     \int     dq     \prod_{a=1}^M     d\mu^a
d\overline{\Pi}_a d\Pi^a (\Upsilon_1(q,i\mu,\Pi,\overline{\Pi}))^*
\Upsilon_2(q,-i\mu,\Pi,\overline{\Pi}).
\l{12}
$$
The integration and conjugation rules are 
$(\overline{\Pi}_{a_1}...\overline{\Pi}_{a_l} \Pi^{b_1} ... \Pi^{b_s})^* = 
(-1)^s \Pi^{b_s}... \Pi^{b_1} \overline{\Pi}_{a_l} ... \overline{\Pi}_{a_1}$, 
$\int d\overline{\Pi}_a \overline{\Pi}_a =1$, 
$\int d{\Pi}^a {\Pi}^a =1$.
Condition \r{9}     means    that    $\Upsilon$    is    $\mu,    \Pi,
\overline{\Pi}$-independent,
$\Upsilon = \Phi(q)$,
provided  that  the ghost   number   of
$\Upsilon$ is  zero  (this  is  a  usual  assumption of gauge theories
\c{KO}). Since
$$
[\Omega,\rho]_+ = -\lambda^a \Lambda_a + \frac{i}{2} \lambda^a f^b_{ab}
- i\lambda^a \overline{\Pi}_b C^c f^b_{ac} - \overline{\Pi}_a \Pi^a,
\l{12*}
$$
for the simplest abelian case one has
$$
(e^{t[\Omega,\rho]_+}\Upsilon) (q,\lambda,  \Pi,   \overline{\Pi})   =
e^{-t\lambda^a\Lambda_a} \Phi(q) e^{-t\overline{\Pi}_a\Pi^a},
\l{12+}
$$
so that
$$
(\Upsilon, e^{t[\Omega,\rho]_+} \Upsilon) = \int dq
\prod_{a=1}^M     d\mu^a
d\overline{\Pi}_a d\Pi^a
\Phi^*(q) e^{it\mu^a\Lambda_a} e^{-t\overline{\Pi}_a \Pi^a} \Phi(q).
$$
Integration over ghost variables gives us $t^M$, so that
$$
(\Upsilon, e^{t[\Omega,\rho]_+} \Upsilon) = \int dq
\prod_{a=1}^M     d\mu^a t^M \Phi^*(q) e^{it\mu^a\Lambda_a} \Phi(q).
\l{12a}
$$
Integrating over $\mu$, we obtain expression \r{3}.

One can  suggest  then  that the correspondence between states in BFV and
projection operator approaches is the following:  one should take  the
BFV state  to  the  gauge  \r{9} and obtain the state $\Phi(q)$ in the
projection operator approach \r{3}.

{\bf 4.} It is much more convenient to obtain the correspondence in  a
gauge-independent form. Consider the BFV transformation \r{8}. Let
$$
X =   X_{00}  (q,\lambda)  +  X_{01}^a(q,\lambda)  \overline{\Pi}_a  +
X_{10,a}(q,\lambda) \Pi^a + ...
$$
For $\Pi=0$, $\overline{\Pi}=0$, one has
$\Omega X|_{\Pi,\overline{\Pi}=0} = (\Lambda_a - \frac{i}{2}  f^b_{ab})
X^a_{01}(q,\lambda)$.
The wave  function  $\Upsilon  = \Phi(q) + \Omega X$ has the following
form on the surface $\Pi,\overline{\Pi},\lambda=0$:
$$
\Upsilon(q,0,0,0) =  \Phi(q)  +  (\Lambda_a  -  \frac{i}{2}  f^b_{ab})
X_{01}^a(q,0).
$$
It coincides with $\Phi(q)$ up to a gauge transformation \r{4} for the
abelian case.  Thus,  we find that projection-operator  and  BFV  wave
functions are related as follows:
$$
\Phi(q) = \Upsilon(q,0,0,0),
\l{13}
$$
while the BFV gauge transformation  \r{8}  corresponds  to  the  gauge
transformation \r{4}.

If we  considered  the coordinate ghost representation of the BFV wave
function, where $C^a$ and  $\overline{C}_a$  are  multiplicators,  the
coefficient of      $C^1...C^M\overline{C}_1...\overline{C}_M$      at
$\lambda=0$ would play a role of the function $\Phi$.

{\bf 5.}  Prescription  \r{13}  can   be   also   justified   in   the
\c{KO,Razumov} approach.  For  the  abelian  case,  one introduces the
creation and annihilation operators
$A_a^{\pm} = \frac{1}{\sqrt{2}} [\pi_a \pm iM_a{}^b \Lambda_b]$
for some Hermitian real positively definite nondegenerate matrix  $M$,
shows that it is possible to perform such a gauge transformation \r{8}
that after it
$$
A_a^- \Upsilon = 0.
\l{14}
$$
It follows from the BFV condition that
$[\frac{\partial}{\partial \overline{\Pi}_a}  + M_b{}^a \Pi^b] \Upsilon
=0$ and
$\Upsilon(q,\lambda,\Pi,\overline{\Pi}) =    \exp    [-\overline{\Pi}_a
M_b{}^a \Pi^b] \Upsilon_0(q,\lambda)$.
Condition \r{14} implies that
$\Upsilon_0(q,\lambda) = \exp [\lambda^a M_a{}^b \Lambda_b] \Phi(q)$,
where $\Phi(q)$ is of the form \r{13}, so that 
$$
\Upsilon(q,\lambda,\Pi,\overline{\Pi}) =    \exp    [-\overline{\Pi}_a
M_b{}^a \Pi^b]  \exp [\lambda^a M_a{}^b \Lambda_b] \Phi(q).
\l{razumov}
$$
Making use of  formula  \r{12}
for the inner product, we find
$$
(\Upsilon,\Upsilon) =   \int   dq   \prod_{a=1}^M   d\mu^a   \Phi^*(q)
e^{- 2i\mu^a M_a{}^b  \Lambda_b}    \Phi(q)     \int     \prod_{a=1}^M
d\overline{\Pi}_a d\Pi^a \exp [-2\overline{\Pi}_a M_b{}^a \Pi^b].
$$
Integration over Grassmannian variables gives us the factor $\det  2M$
which  is  involved  to  the  integration  measure  after substitution
$2\mu^a M_a{}^b = \tilde{\mu}^b$. We obtain formula \r{3}.

{\bf 6.}  Starting  from  formulas  \r{3}  and  \r{10},  let us try to
investigate their range of validity and modify them for  the  case  of
discrete spectrum of $\Lambda_a$.

{\it Example 1.} Let $M=1$, $q=x$, $\Lambda = - i\partial/\partial x$.
The formulas     \r{3}     and     \r{12a}     take      the      form
$|\int_{-\infty}^{+\infty} dx   \Phi(x)|^2$.  This  expression  is
well-defined.

{\it Example   2.}   Let   $M=1$,   $q=\varphi$,    $\Lambda    =    -
i\partial/\partial \varphi$,   $\varphi   \in   (0,2\pi)$,   the  wave
functions be periodic with respect to $\varphi$, $\Phi(\varphi+2\pi) =
\Phi(\varphi)$. Formula \r{12a} takes then the form:
$$
\int_0^{2\pi} \int_{-\infty}^{+\infty}    d\mu    t    \Phi^*(\varphi)
\Phi(\varphi +  \mu  t)  =  \int_0^{2\pi}   d\varphi   \Phi^*(\varphi)
\int_{-\infty}^{+\infty} dy \Phi(y).
$$
However, this  integral  is divergent.  To avoid this difficulty,  one
should perform an  integration  over  the  period  only,  i.e.  $\mu\in
(0,2\pi/t)$. We  obtain  then  the  formula  $|\int_0^{2\pi}  d\varphi
\Phi(\varphi)|^2$ for the inner  product  which  is  a  basis  of  the
projection operator  quantization  \c{proj}.  Thus,  we  see  that the
topology of the group  should  certainly  be  taken  into  account  in
defining the inner product (cf.  \c{DSS}):  formula \r{10}  should  be
corrected  in  such  a  way  that  integration over $\mu$ in eq.\r{12}
should be performed over the finite interval.

{\bf 7.}  Let  us  modify  formula \r{3} for the nonabelian case.  Let
$L_a$, $a=\overline{1,M}$ be generators of the Lie  algebra  with  the
following commutation relations $[L_a,L_b]=if_{ab}^cL_c$. Consider the
corresponding Lie group $G$  and  the  exponential  mapping  $\mu^aL_a
\mapsto \exp(i\mu^aL_a)$.   The   operators   $\Lambda_a$   perform  a
representation of the Lie  algebra,  so  that  $\exp(i\mu^a\Lambda_a)$
will perform   a   representation   of   group  $T(\exp(i\mu^aL_a))  =
\exp(i\mu^a\Lambda_a)$. By  $Ad   (L_a)$   we   denote   the   adjoint
representation of  the  Lie  algebra, $(Ad  (L_a)  \rho)^c = if^c_{ab}
\rho^b$, while $Ad\{g\}$ is an adjoint  representation  of  the  group
$(Ad\{g\} \rho)^c  =  (\exp(A))^c_b  \rho^b$  with  $A^c_b  =  - \mu^a
f^c_{ab}$, $g = \exp(i\mu^aL_a)$.  Find a nonabelian analog of formula
\r{12+}. Let us look for it in the following form:
$$
(e^{t[\Omega,\rho]_+}\Upsilon)(q,\lambda,\Pi,\overline{\Pi})         =
e^{- t\lambda^a \tilde{\Lambda}_a} \Phi(q)  e^{\overline{\Pi}_a  B^a{}_b
(\lambda,t) \Pi^b}
$$
with
$$
\tilde{\Lambda}_a = \Lambda_a - \frac{i}{2} f^b_{ab}.
\l{19}
$$
Making use of the relation
$\frac{d}{dt}(e^{t[\Omega,\rho]_+} \Upsilon) = [\Omega,\rho]_+
e^{t[\Omega,\rho]_+} \Upsilon$, we find the following equation for the
matrix $B$, $\dot{B}^b{}_d = - i\lambda^a f^b_{ac} B^c{}_d - 1$,
so that
$B(\lambda,t) = -\int_0^t d\tau Ad\{\exp(- \tau \lambda^aL_a\} \}$.
Therefore,
$$
(\Upsilon, e^{t[\Omega,\rho]_+}  \Upsilon)  =  \int  dq \prod_a d\mu^a
d\overline{\Pi}_a d\Pi^a  \Phi^*(q)    e^{it\mu^a    ({\Lambda}_a    -
\frac{i}{2} f^b_{ab})}\Phi(q)
e^{-\overline{\Pi}_a
\int_0^t d\tau
(Ad \{ \exp(i\tau \mu^cL_c)\})^a_b \Pi^b}.
\l{15}
$$
Integration over fermionic variables gives us the group measure
$$
dg = \det
\int_0^t d\tau (Ad \{ \exp(i\tau \mu^cL_c)\}) \prod_{a=1}^{\mu} d\mu^a,
\qquad g = \exp(it\mu^cL_c)
$$
It happens that it coincides with the  right-invariant  Haar  measure
which has the form (see,  for example, \c{group}) $d_Rg = d\mu J(\mu)$
with $J(\mu) = \det \frac{\delta \rho}{\delta \mu}$ for
$$
\exp(i(\mu^a+\delta\mu^a)L_a) =        \exp(i\delta         \rho^aL_a)
\exp(i\mu^aL_a).
$$
Without loss of generality, consider the case $t=1$. One finds
$$
\delta\rho^aL_a = \int_0^1 d\alpha e^{i\alpha \mu^aL_a} \delta\mu^b L_b
e^{-i\alpha \mu^aL_a}  =  \int_0^1 d\alpha (Ad\{e^{i\alpha \mu^cL_c}\}
\delta\mu)^a L_a,
\l{15a}
$$
so that $dg=d_Rg$. The multiplicator $e^{t\frac{1}{2} \mu^af_{ab}^b}$
can be presented as $(det Ad\{g\})^{-1/2}$.  The inner product  \r{15}
can be rewritten then as an integral over group
$$
\int dq d_Rg (det Ad \{g\})^{-1/2} \Phi^*(q) T(g) \Phi(q)
\l{16}
$$
with $T(e^{i\mu^aL_a}) = e^{i\mu^a\Lambda_a}$. Note that
$d_Rg (det Ad \{g\})^{-1/2} = d_Lg (det Ad \{g\})^{1/2}$.

Analogously to  the  abelian case,  one can propose that each point of
the gauge group should be taken into account  once.  This  means  that
integration over  $\mu$  in  \r{15} should be performed in general
only over some domain.

Contrary to \c{DSS},  choice \r{9},  \r{10aa} of additional conditions
and gauge fermion leads to well-defined inner products \r{16} for the
models of \c{DSS}. Formula \r{16} coincide with the inner product
derived  in  \c{M2}.  For the compact groups,  analogous formulas were
used in \c{proj} and in lattice gauge theories \c{lattice}.

{\bf 8.}  Formulas  \r{4},  \r{5} can be generalized to the nonabelian
case \c{M2}. Two  states  $\Phi_1$  and  $\Phi_2$  are
gauge-equivalent if their difference satisfies the condition
$$
\int d_Rg (det Ad \{g\})^{-1/2} T(g) (\Phi_1-\Phi_2) = 0.
$$
For example, the transformation
$\Phi \to (det Ad \{h\})^{-1/2} T(h) \Phi$
is gauge. One can also consider infinitesimal gauge transformations
$$
\Phi \to \Phi + \tilde{\Lambda}_a X^a.
\l{17}
$$
The Dirac wave function can be defined as
$$
\Psi = \int d_Rg (det Ad \{g\})^{-1/2} T(g)\Phi.
\l{17a}
$$
It obeys the condition
$$
(det Ad \{h\})^{1/2} T(h)\Psi = \Psi
$$
analogously to eq.\r{2} \c{M2}.
It can be also presented in the infinitesimal form
$$
\tilde{\Lambda}_a^+ \Psi \equiv  (\Lambda_a  +  \frac{i}{2}  f_{ab}^b)
\Psi = 0
\l{18}
$$
found in \c{Duval}.

{\bf 9.}  It  follows  from  eqs.\r{15} and \r{16} that the Dirac wave
function can be also presented via the integral over ghost momenta and
Lagrange multipliers,
$$
\Psi(q) =       \prod_a      d\mu^a      d\overline{\Pi}_a      d\Pi^a
(e^{t[\Omega;\rho]_+} \Phi)(q, -i\mu, \Pi, \overline{\Pi}).
$$
Since states  $\Phi$  and  $e^{t[\Omega;\rho]_+}  \Phi$  are  formally
BFV-equivalent, one   can   suggest  the  following  gauge - independent
correspondence between Dirac and BFV states,
$$
\Psi(q) =    \int    \prod_a    d\mu^a    d\overline{\Pi}_a     d\Pi^a
\Upsilon(q,-i\mu, \Pi, \overline{\Pi}).
\l{Dirac}
$$
Gauge equivalent states give us identical Dirac wave functions,  since
the B-charge can be written as a full derivative,
$$
\Omega =  (\Lambda_a  + \frac{i}{2} f^b_{ab}) \frac{\partial}{\partial
\overline{\Pi}_a} -  \frac{i}{2}   f^a_{bc}
\frac{\partial}{\partial \overline{\Pi}_b}
\frac{\partial}{\partial \overline{\Pi}_c}
\overline{\Pi}_a - \frac{\partial}{\partial \lambda^a} \Pi^a,
$$
so that the integral of $\Omega X$ over $\mu$, $\overline{\Pi}$, $\Pi$
vanishes. Furthermore, it follows from the property
$$
\int \prod_a d\mu^a d\overline{\Pi}_a d\Pi^a (\Omega  \overline{\Pi}_a
\Upsilon)(q,-i\mu,\Pi, \overline{\Pi})
$$
and relation  $\Omega  \Upsilon=0$  that eq.\r{18} is indeed satisfied
for definition \r{Dirac}.

Thus, the {\it formal} relationship between Dirac and  BFV  states  is
obtained analogously to \r{13}. However, integration over $\mu$ should be
performed carefully due to topological problems.  Note that conjecture
\r{Dirac} is indeed valid for the state \r{razumov}.

{\bf 10.}  Let us investigate under what conditions the operator $H$ is
an observable. In the BFV approach, an observable $H_B$ should commute
with BRST-BFV  charge \r{6}.  Consider the expansion of $H_B$ in ghost
momenta \c{Henneaux}:
$$
H_B =   H   +   \overline{\Pi}^a   H_{10}^a   +   \Pi^b   H_{01,b}   +
\overline{\Pi}_a\Pi^b H_{11}{}^a{}_b + ...
$$
We are  to  order  the  ghost operators as follows:  the ghost momenta
should be put to the left, the ghosts are put to the right. One has
$$
\Omega H_B  =  C^c  \tilde{\Lambda}_c H + \tilde{\Lambda}_a H^a_{10} +
..., \qquad H_B \Omega = HC^c \tilde{\Lambda}_c + ...
$$
where $...$  are  terms with ghost momenta,  $\tilde{\Lambda}_a$ is of
the form \r{19}.  Therefore,  $H^a_{10}  =
iR_c^a C^c$, so that the term $H$ should obey the following property:
$$
[H; \tilde{\Lambda}_a]  = i\tilde{\Lambda}_c R_a^c.
\l{20}
$$
The operator $H$ should be  identified  with  the  observable  in  the
projection operator  approach.  Notice that it indeed takes equivalent
states to equivalent,  i.e.  $H\Phi  =  \tilde{\Lambda}_a\tilde{X}^a$,
provided that  $\Phi  =  \tilde{\Lambda}_a  X^a$:  it is sufficient to
choose $\tilde{X}^a = HX^a + iR_b^a X^b$.

An important  feature  of  the  physical  observable   is   that   the
corresponding evolution  operator  $e^{-iHt}$  should  be unitary with
respect to the inner product \r{16}. This means that
$$
H^+ \int d_Rg (\det Ad \{g\})^{-1/2} T(g) =
\int d_Rg (\det Ad \{g\})^{-1/2} T(g) H.
\l{21}
$$
Formula \r{21} implies that state $H\Phi$  corresponds  to  the  Dirac
wave function
$$
\int d_Rg (\det Ad \{g\})^{-1/2} T(g) H\Phi = H^+\Psi.
$$
Therefore, it  is  the operator $H^+$ that corresponds to the physical
observable in the Dirac approach.

Let us illustrate condition \r{21} for the closed-algebra  case,  when
$R_a^c=const$ and  the  $B$-extension of the observable $H$ is written
explicitly \c{Henneaux}:
$$
H_B = H + i\overline{\Pi}_b R_c^b C^c.
\l{22}
$$
Eq.\r{21} to be checked can be rewritten as
$$
\int d_Rg H^+ e^{i\mu^a\tilde{\Lambda}_a}
= \int d_Rg e^{i\mu^a\tilde{\Lambda}_a} H.
\l{23}
$$
One has
$$
e^{i\mu^a\tilde{\Lambda}_a} H  e^{-i\mu^a\tilde{\Lambda}_a}  =   H   +
\int_0^1 d\alpha e^{i\alpha\mu^a\tilde{\Lambda}_a}
[i\mu^a\tilde{\Lambda}_a; H]
e^{-i\mu^a\tilde{\Lambda}_a} =
H +
\int_0^1 d\alpha e^{i\alpha\mu^a\tilde{\Lambda}_a}
\mu^a R_a^b \tilde{\Lambda}_b
e^{-i\mu^a\tilde{\Lambda}_a}.
$$
Since the commutation relations between generators $\tilde{\Lambda}_a$
coincide with    \r{1},    $[\tilde{\Lambda}_a,\tilde{\Lambda}_b]    =
if^c_{ab} \tilde{\Lambda}_c$, it follows from eq.\r{15a} that
$$
e^{i\mu^a\tilde{\Lambda}_a} H  e^{-i\mu^a\tilde{\Lambda}_a}  =   H   +
\frac{1}{i} \frac{d}{d\tau}|_{\tau=0}
e^{i(\mu^a + \tau \mu^b R_b^a) \tilde{\Lambda}_a}
e^{-i \mu^a \tilde{\Lambda}_a}.
$$
Eq. \r{23} is taken then to the form
$\int d_Rg (H^+ - H) e^{i\mu^a\tilde{\Lambda}_a}
= \int d_Rg  \frac{1}{i}  \frac{d}{d\tau}|_{\tau=0}
e^{i(\mu^a + \tau \mu^b R_b^a) \tilde{\Lambda}_a}$,
so that
$$
H=H^+ - iR_b^b
\l{24}
$$
Condition \r{24}  is  a relationship between observables $H$ and $H^+$
in the projection operator and Dirac approaches.  We see that this  is
in agreement with the condition $H_B^+=H_B$.

{\bf 11.} Thus, the correspondence \r{13}, \r{Dirac}
between states in different approaches
to quantize the constrained systems is found.  The inner products used
in different  methods  are modified and generalized.  The relationship
\r{22}, \r{24} between  observables  $H_B$,  $H$  and  $H^+$  in  BFV,
projection operator and Dirac approaches is also found.
Note also that the obtained results can be generalized to the  case
of open algebras \c{Shv1}.

The author  is  indebted  to  Kh.S.Nirov,  D.Marolf,  V.A.Rubakov  and
T.Strobl for helpful discussions. This work was supported by
the Russian Foundation for Basic Research, project 01-01-06251.

\end{document}